\documentclass[a4paper,11pt]{article}
\newcommand{\unit}[2]{#1 \, {\rm #2}}
\usepackage{subfigure}
\pdfoutput=1 

\usepackage{jinstpub} 

\title{\boldmath Radon background in liquid xenon detectors}

\author[a]{Natascha Rupp}

\affiliation[a]{Max-Planck-Institut f\"ur Kernphysik, 69117 Heidelberg, Germany}

\emailAdd{natascha.rupp@mpi-hd.mpg.de}

\abstract{The radioactive daughters isotope of $^{222}$Rn are one of the highest risk contaminants in liquid xenon detectors aiming for a small signal rate. The noble gas is permanently emanated from the detector surfaces and mixed with the xenon target.  Because of its long half-life  $^{222}$Rn is homogeneously distributed in the target and its subsequent decays can mimic signal events. Since no shielding is possible this background source can be the dominant one in future large scale experiments. 
This article provides an overview of strategies used to mitigate this source of background by means of material selection and on-line radon removal techniques.}


\proceeding{LIDINE 2017: LIght Detection In Noble Elements\\
  22-24 September 2017\\
  SLAC National Accelerator Laboratory}

\begin{document}
\maketitle
\flushbottom
\section{Radon as source of background}
\begin{figure}[htbp]
    \centering
    \includegraphics[width=60mm]{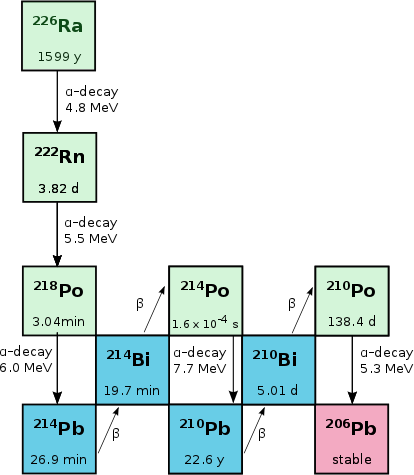}
    \caption{$^{222}$Rn decay chain. }
    \label{fig:chain}
\end{figure}

The radioactive daughter isotopes of radon form an important background source in many low-event rate experiments. Among those neutrino and dark matter experiments rank and we will focus on the ones using liquid xenon as detection medium (for example \cite{internal_bg} - \cite{next_RnBG}). Radon is created by the decay of radium that, being part of the primordial decay chains of thorium and uranium, is present in many materials used for detector construction. Once created, radon can emanate from the material into the xenon reservoir either driven by the recoil energy released in the decay or by diffusion. In the latter case, also radon atoms from deeper material layers can reach the surface. No stable isotopes of radon exist in nature and the one having by far the longest half life is  $^{222}$Rn with T$_{1/2} =  \unit{3.8}{days}$ - sufficient time to travel long distances before its decay. This long half-life in combination with the nonexistence of radon shielding possibilities could turn radon into the dominant background source in future large scale xenon detectors. In Fig.  \ref{fig:chain} the decay chain of  $^{222}$Rn is shown, on which we will concentrate in the following discussion. Usually the $\beta$ decaying isotopes within the chain are the crucial background contributors to dark matter search and neutrinoless double beta decay experiments. Thereby the decays subsequent to  $^{214}$Po are only present if the detector material is exposed to a large radon concentration and a significant amount of radon daughters plate out on the surface. Otherwise they are suppressed due to the long half-life of $^{210}$Pb. Those large radon concentrations can be typically reached in air and not from the emanation of the detector material itself. The decay of the out-plated radon daughters at the edges of the detector can generate background events, whereby also the $\alpha$ decaying isotopes can contribute as their signal can be degraded. If we only consider decays up to $^{210}$Po, the two remaining  $\beta$ decaying isotopes are $^{214}$Bi and $^{214}$Pb. The latter one is most important in dark matter search \cite{bruenner_diplom}. Because of its lower Q-value in comparison with $^{214}$Bi, its decay rate is the highest in the typical region of interest ($ <  \unit{70}{keV}$). In the search for neutrinoless double $\beta$ decay the $^{214}$Bi decay is most important, since it has a  $\gamma$ line (2448 keV) near the $\beta\beta$ Q-value of  $^{136}$Xe \cite{exo_RnBG}. By making use of the $^{214}$Bi - $^{214}$Po coincidence a large part (typically >50$\%$  \cite{internal_bg} \cite{exo_RnBG}) of the decays can be tagged if the $^{214}$Bi decay happens in a sensitive detection region. The $\alpha$  decaying isotopes within the chain usually don't contribute to the radon induced background because their decay energies are in the MeV region and mono energetic, which makes them easy to identify in the data analysis. By measuring their rates one can set limits on the $\beta$ decay rates, that can differ among each other, even though the decay chain is in equilibrium \cite{internal_bg} - \cite{panda_ana}. The different daughter isotopes of $^{222}$Rn can stay positively charged after the decay and hence accumulate at the cathode or detector edges, whereby they are removed from the sensitive region \cite{exo_ion} \cite{internal_bg}. 

\section{Radon screening}
One method to mitigate the radon induce background is a careful material selection of radio-pure materials. A common way is $\gamma$ spectroscopy to measures the  $^{214}$Bi and $^{214}$Pb decay rates, from which the total expected $^{222}$Rn rate can be inferred. However, this expected rate can differ from the actual emanating $^{222}$Rn rate because the radon atoms might be inhomogeneously distributed in the material. Additionally, the emanation rate depends on the radon diffusion length in the particular material. Therefore, it is more favorable to perform an actual $^{222}$Rn emanation measurement \cite{seb} \cite{Natascha} \cite{bruenner_doktor}.  In a first step the sample is enclosed in an evacuated vessel that is then filled with radon cleaned carrier gas, typically helium. After some days, it is extracted together with the radon atoms that emanated from the sample. In a next step the carrier gas and impurities, that also emanate from the sample (e.g. organic outgassing) are removed from the radon atoms by means of gas chromatography. In the end the cleaned sample is filled into an alpha detector together with counting gas (typically ArCH$_{4}$). By measuring the $\alpha$ decays one can infer the radon emanation rate of the sample. Miniaturized proportional counter can be used as alpha detector featuring a low background of $\sim\unit{1}{count/day}$ and a high sensitivity of $ \unit{20}{\mu Bq}$ \cite{counter_paper}. Another possibility are electrostatic radon monitors that measure the decay of $^{222}$Rn daughters with a PIN diode. Radon monitors have the advantage that one can measure big gas samples ($\mathcal{O}(100 \ l)$) on-line.
\section{Radon removal techniques}
Once the detector is constructed the radon emanation rate is set. A further radon reduction could be achieved by so-called radon removal system integrated in the xenon recirculation loop that continuously separates the xenon from radon atoms. The radon atoms are retained until their disintegration and only xenon can pass and be flushed back into the detector. However, the similar physical properties of xenon and radon hamper their separation. Furthermore, depending on the exact location of the radon sources within the detector system large recirculation fluxes are needed to flush the radon atoms out of the sensitive volume before they decay. Those high recirculation fluxes of $\mathcal{O}( \unit{100}{STPM})$  imply further technical challenges like a sufficient recirculation pump performance.  Adsorption and cryogenic distillation are two promising approaches for radon removal as explained in the following. 
\subsection{Adsorption}
Radon removal by adsorption has been mainly investigated by several groups \cite{xmass_adsoption} \cite{bruenner_diplom} \cite{talk_LZ}. (Physical) adsorption is based on the van-der-Waals forces and polarizability of the atoms and the mentioned groups studied activated carbons as adsorbents as they have a large surface area and a micropore volume. The XMASS collaboration proved that radon atoms can be retained in cooled activated carbon and thereby sufficiently removed from xenon. It was also shown that the radon self-emanation of the adsorbent itself ultimately limits the radon reduction power. The adsorption capacity of different adsorbents was systematically investigated in \cite{bruenner_diplom}. Besides the potential high radon self-emanation of the adsorbent one has to consider the rather high adsorption capacity of xenon. Measurements in \cite{seb} showed that the mass of adsorbed xenon per charcoal mass is about $ \unit{1.3}{g/g}$ at a pressure of $ \unit{1}{bar}$  and a temperature of $-80^\circ\text{C}$ . Therefore a larger amount of additional xenon would be required in the detector because a fraction will be always adsorbed. 
\subsection{Cryogenic distillation}
Cryogenic distillation is already successfully applied in the separation of xenon from $^{85}$Kr, which is another important internal background source in liquid xenon experiments \cite{xenon_krypton} - \cite{wang}. In the distillation process, the more volatile component of a binary liquid (e.g. a xenon and radon mixture) gets enriched in the gas blanket above the liquid surface. In case of a xenon/radon mixture, xenon is the more volatile component and enriched in the gas phase. In \cite{bruenner_boiloff} a $^{222}$Rn reduction factor of R >  4 is measured in the boil-off xenon gas with respect to to the radon enriched liquid phase. This result can be understood as a single-stage distillation process. In the framework of the XENON100 experiment, the $^{222}$Rn reduction power was increased by operating an on-line multi-stage distillation column integrated in the detector recirculation system \cite{xenon_rn_dist}. The column was originally developed for the krypton-xenon separation in XENON100 and run in reversed mode to separate radon from xenon. A radon reduction factor of R > 27  ($\unit{95}{\%}$ C.L.) was determined for the distillation column. First tests in the successor experiment XENON1T showed a $^{222}$Rn reduction by $\sim \unit{20}{\%}$ by looping only a small fraction of the recirculation flow \cite{first_result} through a distillation column. Further studies are ongoing for the construction of a dedicated radon distillation column that can also handle higher recirculation flows.


\begin{thebibliography}{99}

\bibitem{internal_bg}
XENON Collaboration, \emph{Intrinsic backgrounds from Rn and Kr in the XENON100 experiment},  arXiv:1708.03617 (2017).

\bibitem{lux_ana}
LUX Collaboration, \emph{Radon-Related Backgrounds in the LUX Dark Matter Search},  Physics Procedia 61 658 665 (2015).

\bibitem{panda_ana}
PandaX Collaboration, \emph{Krypton and radon background in the PandaX-I dark matter experiment},  arXiv:1701.07307v1 (2017).

\bibitem{exo_RnBG}
EXO Collaboration, \emph{Investigation of radioactivity-induced backgrounds in EXO-200}, Phys. Rev.  C 92, 015503 (2015).

\bibitem{next_RnBG}
NEXT Collaboration, \emph{Radon and material radiopurity assessment for the NEXT double beta decay experiment}, arXiv:1505.07052 (2015).

\bibitem{bruenner_diplom}
S. A. Bruenner, \emph{Study of radon adsorption on activated carbon for a purification system in XENON1T}, Master thesis, University of Graz, (2013).

\bibitem{xenon_physics_reach}
XENON Collaboration, \emph{Physics reach of the XENON1T dark matter experiment}, 	arxiv: 1512.07501 (2015).

\bibitem{exo_ion}
EXO Collaboration, \emph{Measurements of the ion fraction and mobility of alpha and beta decay products in liquid xenon using EXO-200}, arXiv:1506.00317v2 (2017).

\bibitem{seb}
S. Lindemann. \emph{Intrinsic 85Kr and 222Rn Backgrounds in the XENON DarkMatter Search} PhD thesis, University of Heidelberg, (2013).

\bibitem{Natascha}
N. M. Rupp \emph{On the detection of 222Rn with miniaturized propotional counters: background, sensitivity studies and results for XENON1T}, Master thesis, University of Heidelberg, (2015).

\bibitem{bruenner_doktor}
S. A. Bruenner, \emph{Mitigation of 222Rn induced background in the XENON1T dark matter experiment}, Phd thesis, University of Heidelberg (2017).

\bibitem{counter_paper} 
G. Zuzel, H. Simgen \emph{High sensitivity radon emanation measurements}, Appl. Radiat. Isot. 67(5):889-93 (2009).

\bibitem{xmass_adsoption} 
XMASS Collaboration, \emph{Radon removal from gaseous xenon with activated charcoal}, NIMA A 661 (2012) 50 57 (2012).

\bibitem{talk_LZ} 
LZ Collaboration, \emph{Prototype radon removal system for the LZ experiment}, Talk at LRT (2017).

\bibitem{xenon_krypton}
XENON Collaboration, \emph{Removing krypton from xenon by cryogenic distillation to the ppq level}, 
Eur.Phys.J. C 77 (2017).

\bibitem{abe_krypton}
XMASS Collaboration, \emph{Distillation of Liquid Xenon to Remove Krypton}, Eur.Phys. J. C 77: 275 (2009).

\bibitem{wang}
Z. Wang et al., \emph{Large scale xenon purification using cryogenic distillation for dark matter detectors}, Rev. Sci. Instrum. 85, 015116 (2014).

\bibitem{bruenner_boiloff}
S. A. Bruenner, D. Cichon, S. Lindemann, T. Marrod\'an Undagoitia, and H. Simgen,  \emph{Radon depletion in xenon boil-off gas}, Eur. Phys. J., C77(3):143, (2017).

\bibitem{xenon_rn_dist}
XENON Collaboration, \emph{Online 222Rn removal by cryogenic distillation in the XENON100 experiment},  arXiv:1702.06942 (2017).

\bibitem{first_result}
XENON Collaboration, \emph{First Dark Matter Search Results from the XENON1T Experiment}, 	PRL 119, 181301 (2017). 



\end{thebibliography}
\end{document}